\documentclass[prd,twocolumn,nofootinbib,aps,floats,floatfix,amsmath,amssymb,secnumarabic]{revtex4} %

\usepackage{graphicx}
\usepackage[usenames,dvipsnames]{color}
\usepackage{amsmath,amssymb}
\usepackage{slashed}
\usepackage{verbatim}
\usepackage[colorlinks,citecolor=blue]{hyperref}
\usepackage{ulem}

\def\sfrac#1#2{{\textstyle{#1\over #2}}}

\newcommand{\be}{\begin{equation}}
\newcommand{\ee}{\end{equation}}
\newcommand{\ba}{\begin{array}}
\newcommand{\ea}{\end{array}}
\newcommand{\bea}{\begin{eqnarray}}
\newcommand{\eea}{\end{eqnarray}}
\newcommand{\sss}{\scriptscriptstyle}

\begin{document}
\title{Completing constrained flavor violation: lepton masses,
neutrinos and leptogenesis}
\author{James M.\ Cline\footnote{jcline@physics.mcgill.ca}}
\affiliation{Department of Physics, McGill University,
3600 Rue University, Montr\'eal, Qu\'ebec, Canada H3A 2T8}
\affiliation{Niels Bohr International Academy and Discovery Center,
Niels Bohr Institute, University of Copenhagen,
Blegdamsvej 17, DK-2100 Copenhagen \O, Denmark}
\author{Alfonso Diaz-Furlong\footnote{adiazfurlong@yahoo.com}}
\affiliation{Facultad de Psicologia, Benemerita Universidad Autonoma de Puebla, 4 sur, Centro Historico,
Puebla, Pue., Mexico, C.P. 72000}
\affiliation{Department of Physics, McGill University,
3600 Rue University, Montr\'eal, Qu\'ebec, Canada H3A 2T8}
\author{Jing Ren\footnote{jren@physics.utoronto.ca}}
\affiliation{Department of Physics,
University of Toronto,
Toronto, Ontario, Canada M5S1A7}

\begin{abstract}
Constrained flavor violation is a recent proposal for predicting
the down-quark Yukawa matrix in terms of those for up quarks and
charged leptons.
We study the viability of CFV with respect to its predictions for
the lepton mass ratios, showing that this remains a challenge, and
suggest some possible means for improving  this shortcoming. We then
extend CFV to include neutrinos, and show that it leads to interesting
predictions for hierachical heavy neutrinos, and leptogenesis
dominated by decays of the second heaviest one (``N2 leptogenesis''),
as well as the possibility of low-scale leptoquark-mediated exotic
decays.
\end{abstract}

\maketitle

\section{Introduction}

\begin{figure}[t]
\includegraphics[width=0.5\columnwidth]{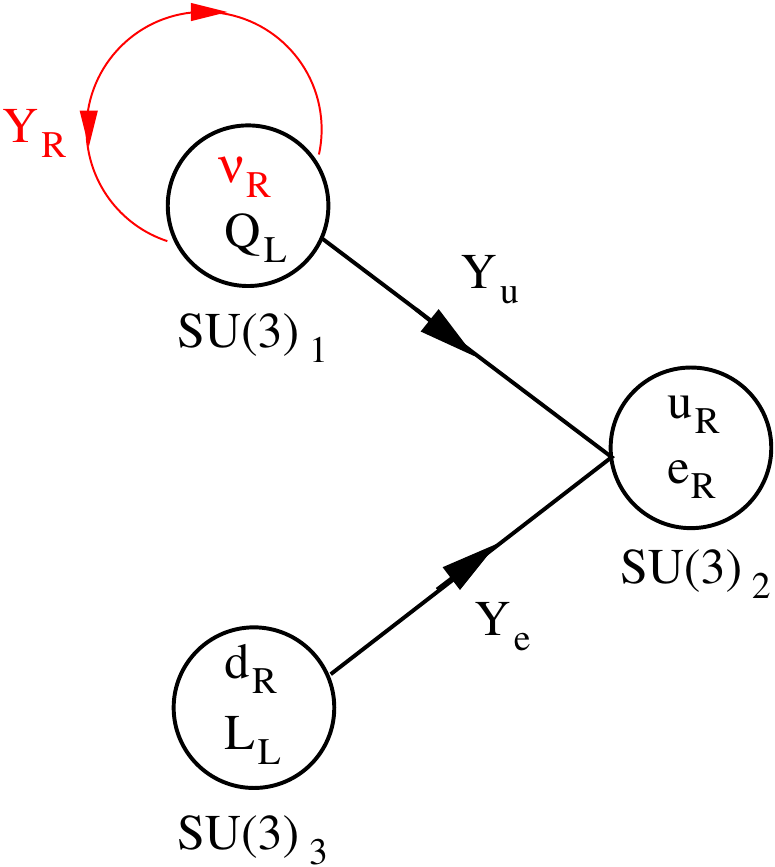}
\caption{Moose diagram for the model, including our proposal for
how the right-handed
neutrinos transform.
}
\label{moose}
\end{figure}

\begin{table*}[t]
\begin{tabular}{|c|c|c|c|c|c|c|c|c|c|c|c|c||c|c|}
 \hline
model &$\lambda$ &  $A$  &  $\rho$ &  $\eta$ & $m_{ud}$ & $m_u/m_d$ & $m_u$ & $m_c$ & $m_t$ & $m_d$  & $m_s$ & $m_b$ &
$m_e/m_\tau$ &  $m_\mu/m_\tau$ \\
& & & & & (MeV) & & (MeV) & (GeV) & (GeV) & (MeV) & (MeV) & (GeV) & $(10^{-4})$ & $(10^{-2})$ \\
\hline
1& 0.2193& 0.91& $-0.062$& 0.371& 1.13 &0.95& 1.10& 0.119& 78.2& 1.16& 13.3& 1.10&  2.784&  5.889\\
2& 0.2195& 0.91& $\phantom{-}$0.030& 0.344& 1.14 &0.93& 1.10& 0.124& 85.2& 1.18& 16.2& 1.09&  2.787&  5.893\\
3& 0.2192& 0.90& $-0.054$& 0.374& 1.31 &0.82& 1.18& 0.124& 84.2& 1.44& 24.0& 1.02&  2.785&  5.863\\
4& 0.2211& 0.90& $-0.048$& 0.341& 1.20 &0.93& 1.16& 0.134& 79.6& 1.24& 20.1& 1.03&  2.783&  5.894\\
5& 0.2204& 0.88& $-0.033$& 0.346& 1.35 &0.97& 1.33& 0.130& 78.6& 1.37& 31.3& 1.09&  2.793&  5.877\\
\hline
SM & 0.2205 & 0.8797 & 0.0 & 0.371 & 0.81 & 0.46 & 0.48 & 0.235 &  74.0 & 1.14 & 22 & 1.0 & $2.786$
& 5.882 \\
error &  0.0006 & 0.024 & $0.021$ & 0.013 & $0.25$ & $-$ & 0.18 &  0.04 & 3.85 & 0.5 & 6.5 & 0.04 & $-$ & $-$ \\
\hline
\end{tabular}
\caption{Upper rows: best-fit models from random scan over GUT-scale parameters to lepton mass ratios,
including KM variations of $m_u/m_d \in [0,1]$.  $\lambda,A,\rho,\eta$ are Wolfenstein parameters for
the CKM matrix.  Lower rows: standard model central values and errors used for the scan.}
\label{tab:models}
\end{table*}

Minimal flavor violation (MFV) \cite{Chivukula:1987py,D'Ambrosio:2002ex} has been an extremely useful
framework for parametrizing effects of new physics in which flavor
symmetry is assumed to be spontaneously broken.  Recently
ref.\ \cite{Appelquist:2015mga} proposed a more predictive version of
MFV in which there are only two fundamental Yukawa
matrices, $Y_e$ and $Y_u$ (for charged leptons and up-type quarks),
while the third one $Y_d$ (for down-type quarks) is predicted to be
the product,
\be
	Y_d = \eta\, Y_u Y_e^\dagger
\label{Yd}
\ee
at tree level, where $\eta\cong 10^3$ to fit the observed lepton
masses.  The structure (\ref{Yd}) is a consequence of a spontaneously
broken flavor symmetry SU(3)$_1\times$SU(3)$_2\times$SU(3)$_3$, under
which the SM fields $Q,L$ (left-handed doublets) and $u,d,e$
(right-handed singlets) transform, as shown in fig.\
\ref{moose}\footnote{{\color{black}{The fermion fields transform as fundamental of each $SU(3)_i$, while the spurion fields transform as: $Y_u\to U_1 Y_u U_2^\dag$, $Y_e\to U_3 Y_e U_2^\dag$ with $U_i\in SU(3)_i$.}}}.  It is argued that the charged lepton mass ratios
are predicted almost correctly in this framework.   Clearly, if it
were possible to reduce the number of free parameters in the
fundamental theory by the elimination of $Y_d$, this could have
profound implications for the ultimate explanation of the flavor
structure of the standard model.  The authors dub this scenario
``constrained flavor breaking;'' here we call it
``constrained flavor violation''(CFV) in analogy to MFV.

Our goal in this paper is two-fold.  First we examine the prediction
of the lepton mass ratios more closely, since ref.\
\cite{Appelquist:2015mga} found that $m_\mu/m_\tau$
is too low except in a few models based upon large fluctuations of
the  quark masses and mixings away from their measured central values.
We will show that this is not an easy problem to solve, and that
without any additional caveats it is more severe than suggested by
ref.\ \cite{Appelquist:2015mga}.  We find it necessary to
implement the predictions at the GUT scale rather than at $m_Z$,
and to suppose that there is some means for altering the prediction for
the up-to-down quark mass ratio at this scale relative to the
weak scale. In particular we suggest that
model-dependent threshold corrections, coming from integrating out
heavy flavon scalars, could improve the situation.  (Recently
ref.\ \cite{Guadagnoli:2015nra} showed that alternate possible relations
between the Yukawa matrices could also give some modest improvement in the
predicted lepton mass ratios.)

Our second goal is to make a proposal for how to bring neutrinos into this
framework, and explore its consequences.  As was pointed out in ref.\
\cite{Appelquist:2015mga}, one of the virtues of CFV is that it
predicts a very nonhierarchical form for $Y_e$, in the special
basis where $Y_d$ is diagonal, which could naturally explain the
large mixing angles of the neutrino sector, even if the neutrino
Yukawa matrix $Y_\nu$ is close to being diagonal.  We will demonstrate
this explicitly by proposing that (like $Y_d$) $Y_\nu$ is not a
fundmental input but is rather also determined by $Y_u$ and $Y_e$
in a manner similar to eq.\ (\ref{Yd}).

We will pursue the consequences of our proposal for leptogenesis,
showing that it leads to a very hierarchical structure for the
heavy right-handed neutrino masses, and naturally fulfilling conditions
where decays of the intermediate mass states $N_2$ tend to give the
dominant contribution to the baryon asymmetry.  We find that a random
scan of parameter space, constrained by the observed charged lepton mass
ratios, can result in a sufficient baryon asymmetry for a large fraction
of generated models.

We also briefly consider some possible consequences of CFV for lepton flavor
violation that were not previously discussed, that are suggestive of
leptoquark-mediated interactions at a scales ranging from $\gtrsim
2-260\,$TeV.

\begin{figure*}[t]
\centerline{\includegraphics[width=1.8\columnwidth]{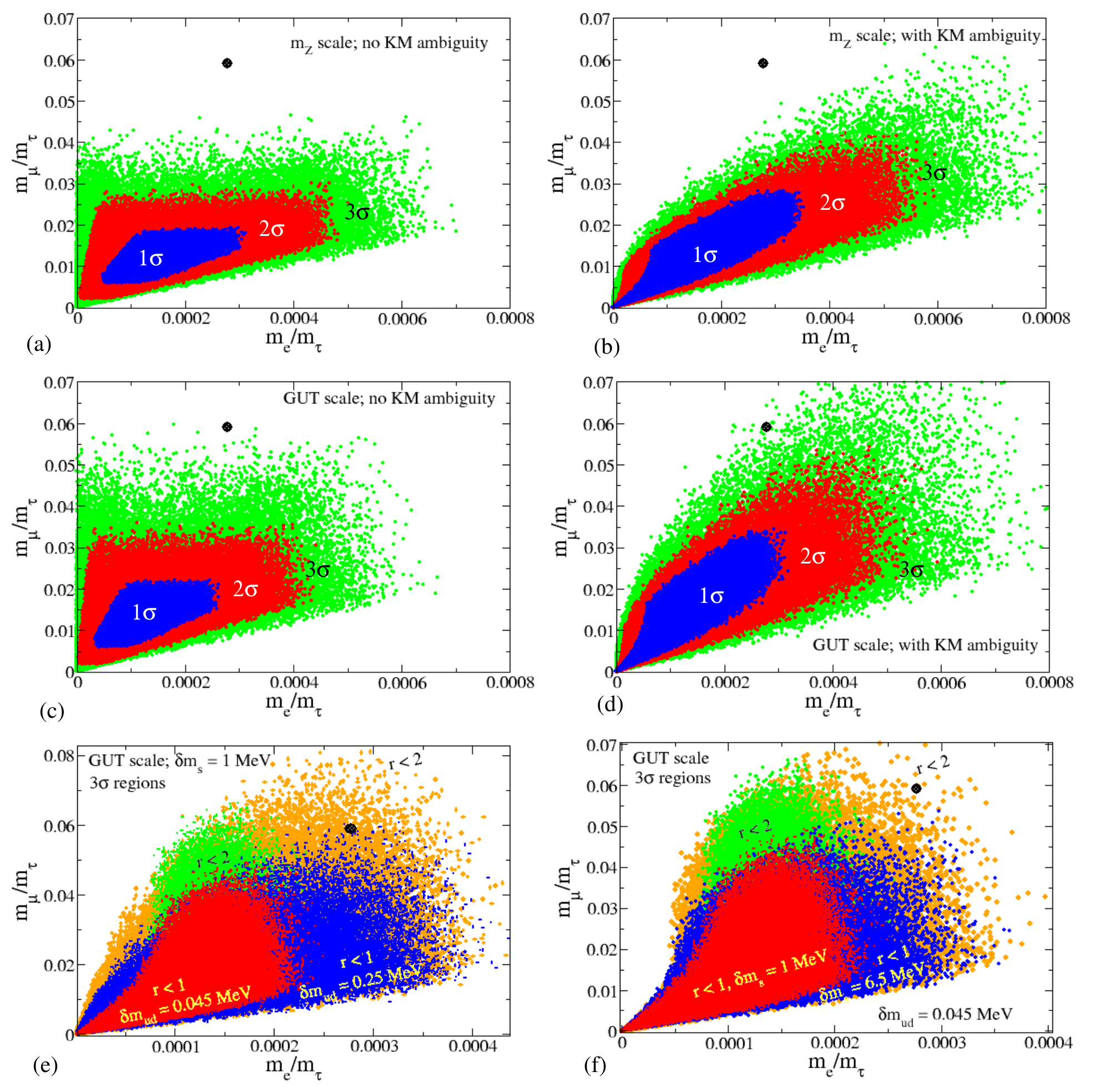}}
\caption{Scatter plots for the lepton mass ratios $m_\mu/m_\tau$
versus $m_e/m_\tau$, with 1-, 2- and 3-$\sigma$ variations of the
quark mass and mixing input parameters around their measured central
values. Shaded dot shows experimental value. In the upper two rows,
top (middle) row is for inputs at the $m_Z$ (GUT) scale;
left (right) column is without (with) the Kaplan-Manohar (KM) ambiguity in
the light quark masses.   Third row shows $3$-$\sigma$ allowed regions
for different assumptions about the experimental errors on $m_{ud} =
(m_u+m_d)/2$ and $m_s$, as well as the range of $r = m_u/m_d$ probed
by KM transformations.}
\label{scatter-panel}
\end{figure*}

\section{Lepton mass ratios}
\label{LMR}

\begin{figure*}[t]
\includegraphics[width=2\columnwidth]{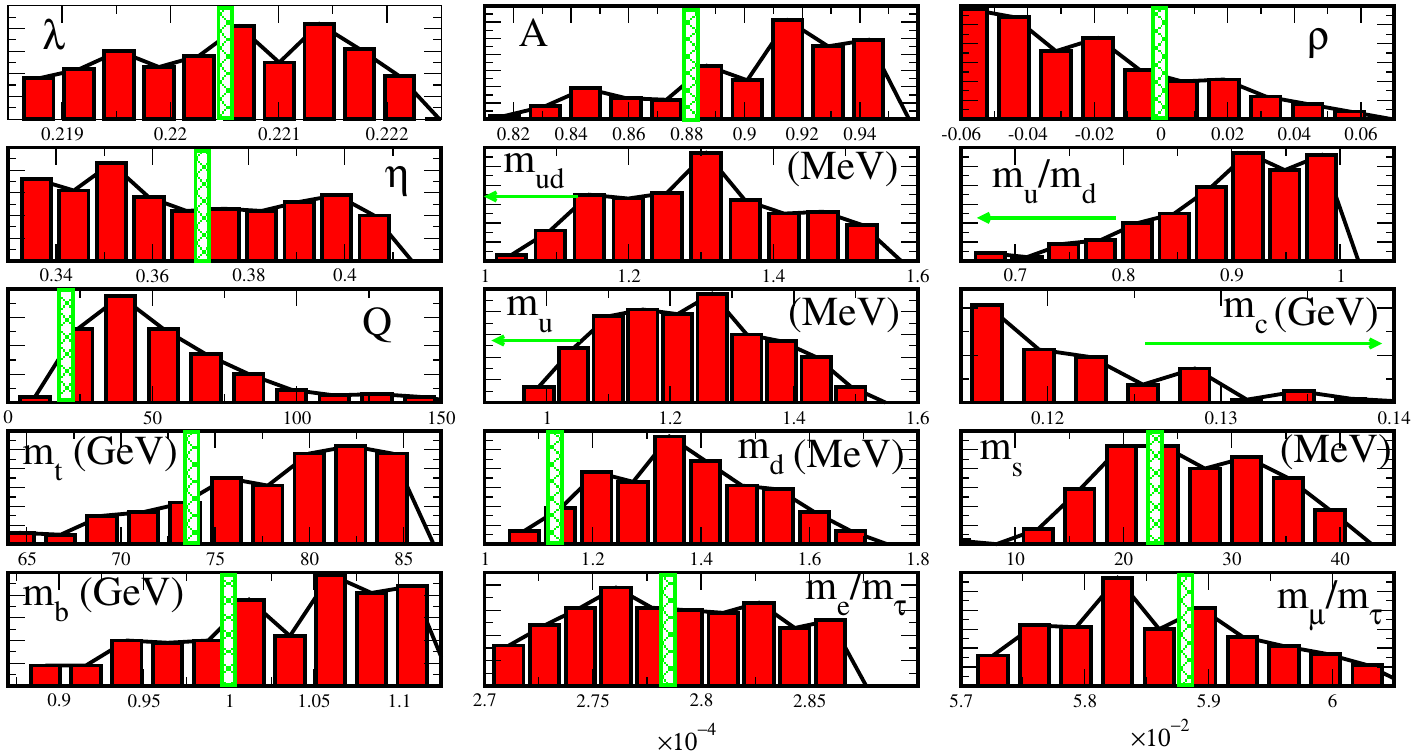}
\caption{Distributions of quark parameters at the GUT scale,
consistent with observed lepton mass ratios, assuming that
${\rm max}|h_{ij}|=1$.  Hatched bars indicate
standard model values, if present within the plotted range; if not,
arrows indicate direction of the SM central value.  $\lambda$, $A$,
$\eta$, $\rho$ are Wolfenstein CKM parameters, $m_{ud} = (m_u+m_d)/2$
and $Q^2 = (m_s^2-m_{ud}^2)/(m_d^2-m_u^2)$.
}
\label{hist}
\end{figure*}

In this section we reexamine the prediction of CFV for ratios of the
charged lepton masses.   This is a fundamental test since these
arise directly from the prediction (\ref{Yd}), by solving for $Y_e$,
\be
	Y_e = \eta^{-1}(Y_u^{-1}Y_d)^\dagger =
	\eta^{-1}{\rm diag}(Y_d) V^\dagger_{\rm\sss CKM}\,
	{\rm diag}(Y_u)^{-1}
\label{Ye_pred}
\ee
where the last expression is written in the basis where $Y_d$ is
diagonal and $Y_u = V^\dagger_{\rm\sss CKM}\, {\rm diag}(Y_u)$.
By inputing the measured quark masses and CKM mixings, within
experimental uncertainties, one can generate $Y_e$ from
(\ref{Ye_pred}), find its eigenvalues, and compute the mass ratios
$m_\mu/m_\tau$, $m_e/m_\tau$, independently of the adjustable
parameter $\eta$.  Ref.\ \cite{Appelquist:2015mga} carries this out
for a large ensemble of randomly generated models, taking $1$-, $2$-
and $3$-$\sigma$ variations in the input parameters; only for a small
fraction near the edge of the $3\sigma$ allowed region is
$m_\mu/m_\tau$ as large as its measured value.

In ref.\ \cite{Appelquist:2015mga}'s implementation, rather generous
ranges are taken for the quark masses at the scale of $m_{Z}$,
whose origin is not explained.
Here we adopt the running quark masses at the scale $m_Z$ along
with uncertainties as given in ref.\ \cite{Xing:2007fb},
and ranges given for the CKM matrix elements by the
Particle Data Group \cite{PDG}, reproduced
in appendix \ref{quark_masses}.  With these inputs, a scan over
$3\times 10^5$ models fails to produce any with $m_\mu/m_\tau >
0.045$ even at 3$\sigma$, whereas the observed value is close to
$0.06$.  This result is plotted in fig.\ \ref{scatter-panel}(a)
(upper left).  The discrepancy is worse than found in ref.\
\cite{Appelquist:2015mga}, due to their larger and unexplained estimates of
the experimental errors.

One might question whether the prediction (\ref{Yd}) is valid at the
scale $m_Z$, whereas the UV flavor physics is expected to come in at a higher
scale.  To assess the effect of going to higher scales, we take
advantage of the running Yukawa couplings (represented as running
masses) calculated in ref.\ \cite{Xing:2007fb} to also test (\ref{Yd})
at the GUT scale, taken to be $2\times 10^{16}$ GeV.  For consistency,
one also needs the CKM parameters at this scale, which we take from
ref.\ \cite{Fusaoka:1998vc}.
The result is
shown in fig.\ \ref{scatter-panel}(c) (middle left), giving
considerable improvement, though the observed lepton mass ratios still
remain at the very edge of the 3$\sigma$ region where the scatter plot
is sparsely populated.

In an attempt to address the shortfall in $m_\mu/m_\tau$, ref.\
\cite{Appelquist:2015mga} makes a parametric estimate $m_\mu/m_\tau
\sim (m_b/m_s)(m_u/m_c)\lambda$ (where $\lambda = \cos\theta_C$ is the
Wolfenstein parameter), suggesting that a larger value
of $m_u/m_s$ could ameliorate this problem.   According to lattice
determinations of the light quark masses, there is no latitude, beyond
the usual error estimates, for increasing $m_u/m_s$ (for a recent
review see \cite{Aoki:2013ldr}).  However, there
are still no direct lattice determinations of this ratio.  Instead,
the up and down quark are always represented by the same field, having
a mass of $m_{ud} = (m_u+m_d)/2$.  Phenomenological input using chiral
perturbation theory (ChPT) is required to estimate the isospin breaking
effects from $m_u\neq m_d$.

Kaplan and Manohar \cite{Kaplan:1986ru} (KM) pointed out that at
second order in ChPT there is an operator $(\det M){\rm
tr}(M^{-1}\Sigma)$
that effectively transforms the quark masses
by
\bea
	m_u &\to& m_u + \alpha\, m_d m_s\nonumber\\
	m_d &\to& m_d + \alpha\, m_u m_s\nonumber\\
	m_s &\to& m_s + \alpha\, m_u m_d
\eea
where $\alpha$ is a parameter of order $1/\Lambda_{\sss QCD}$.
This could shift the apparent quark masses as deduced from ChPT
away from the true values, with a much bigger effect on $m_u$ and
$m_d$ than on $m_s$.  In principle, $\alpha$ can be determined by
comparing enough measured quantities to their second order ChPT
predictions (thus determining all the second order coefficients), and
this procedure would thus resolve the KM ambiguity.  On this basis,
the isospin breaking effects are considered to be well understood and
the ratio
$r\equiv m_u/m_d$ is known to high precision
$0.46\pm 0.02\pm 0.02$ from simulations with $2+1$ flavors (2
denoting degenerate $u$ and $d$, and $1$ denoting $s$).  However this
requires the implicit assumption (not usually stated) that third order
ChPT contributions are negligible.  This assumption can only be
rigorously
tested by doing a full $1+1+1$ lattice simulation, leaving room for
some doubt about the true value of $r$.\footnote{We thank D.B.\ Kaplan
for discussion on this point.}  In fact such simulations have recently
been performed \cite{Borsanyi:2014jba,Horsley:2015eaa}.  This would
seem to close the door on this loophole.  Nevertheless it inspired
us to consider the possibility of allowing $m_u/m_d$ to differ
from its standard value.  Below we will suggest an alternative possible
justification for doing so.

Hence we define
$r\equiv m_u/m_d$ and allow it to vary away from its standard value, while
keeping the errors on $m_{ud}$ and $m_s$ consistent with ref.\
\cite{Xing:2007fb}.  We allow $r$ to vary in the interval $[0,1]$
to obtain the augmented allowed regions shown in fig.\
\ref{scatter-panel}(b,d).  This does not improve the situation for
Yukawa couplings at the $m_Z$ scale, but it does improve it somewhat
at the GUT scale.  Allowing $r$ to vary more widely, $r\in[0,2]$,
can further improve the overlap, as shown in the fig.\
\ref{scatter-panel}(e,f) (bottom row).\footnote{Proton stability could be
consistent with $r>1$ if $r$ runs to smaller values in going from the
GUT to the QCD scale.}
   Parameters of the three best-fit models from this
scan are given in table \ref{tab:models}, along with the central values and
errors for the varied parameters.
 In Fig.~\ref{hist} we show the distributions of
quark mass and mixing parameters resulting from the scan
at the GUT scale.

Before declaring a modest victory however, it should be noted that
many lattice practitioners consider the errors on the light quark
masses like those quoted in \cite{Xing:2007fb,PDG} to
be overestimates that do not reflect the state-of-the-art
lattice results.  In figs.\ \ref{scatter-panel}(c,d), we took the
errors to be $\delta m_{ud} = 0.25$ MeV, $\delta m_s = 6.5$ MeV.
The rare points in our scans that agree with the lepton mass ratios
rely upon large upward fluctuations in $m_{ud}$ and downward
ones in $m_s$.  Figs.\ \ref{scatter-panel}(e,f) demonstrate that
these rare fluctuations are eliminated by taking the smaller estimates
$\delta m_{ud} = 0.045$ MeV and $\delta m_s = 1$ MeV inferred from
the lattice results \cite{Xing:2007fb} after rescaling to account for
the running of the masses to lower values at the GUT scale.  We can
afford to take the tighter error bar on either $m_{ud}$ or $m_s$,
but not both, and still get agreement with the lepton masses.  Taking the
smaller errors on $m_{ud}$ and $m_s$, even with
compensating large values of $r\lesssim 2$, although $m_\mu/m_\tau$ can be large
enough, a new problem arises in that $m_e/m_\tau$ is predicted to be
too small.

We have seen that renormalization effects are quite important in the
interpretation of the CFV prediction (\ref{Ye_pred}).  It is
conceivable that the flavons whose VEVs give rise to the Yukawa
couplings have masses over some range of scales, which could induce
threshold corrections in the running of the Yukawas and perhaps
explain a larger value of $r$ at the GUT scale than at low scales,
avoiding the need for invoking the KM ambiguity. We do not
attempt any such model-building here, but this could
motivate giving further consideration to CFV.  In the following we
suggest an extension of CFV that encompasses the neutrino sector.

\begin{figure*}[t]
\includegraphics[width=2\columnwidth]{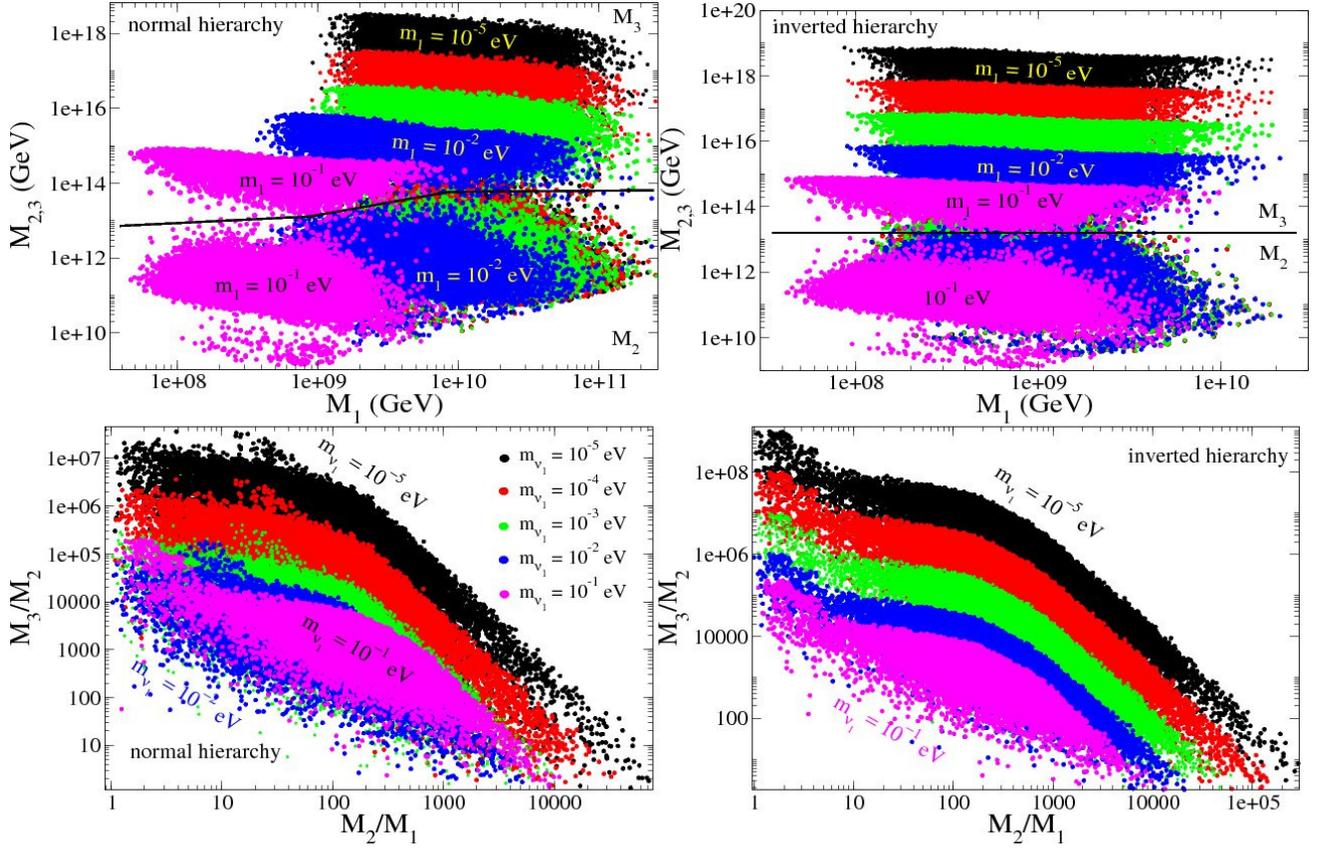}
\caption{Top row: distributions of heavy neutrino masses $M_3$ versus $M_1$
or $M_2$ versus $M_1$ (above and below horizontal line, respectively)
for different choices
of the lightest neutrino mass $m_1= 10^{-5},\,10^{-4},\dots\, 10^{-1}$
eV.  Bottom row: distributions of mass ratios, $M_3/M_2$ versus
$M_2/M_1$.
Left: normal mass hierarchy; right: inverted mass hierarchy.}

\label{RHNmasses}
\end{figure*}

\section{Completing CFV with neutrinos}

To incorporate neutrinos into CFV,
one needs to make some assumption about how the right-handed neutrinos
$\nu_R$ transform under the flavor symmetries.  In CFV,
two species transform nontrivially under each
one of the SU(3) subgroups, with the exception of SU(3)$_1$,
as depicted in fig.\ \ref{moose}.
It is therefore
natural to assign $\nu_R$ to the SU(3)$_1$ node of the moose diagram.
To predict the
neutrino masses and mixings, we must introduce an additional symmetric spurion
field $Y_R$ for the right-handed Majorana neutrino mass matrix, that
transforms in the $\bar 6$ (the symmetric part of $\bar 3\times \bar
3$) representation of SU(3)$_1$. Then, the Lagrangian invariant under flavor symmetry includes following Yukawa interactions and the right-handed Majorana neutrino mass term,
\bea
{\cal L} &=&	-H\bar Q_L Y_d d_R - \tilde H \bar Q_L Y_u u_R
	-H\bar L_L Y_e e_R - \tilde H\bar L_L Y_\nu \nu_R
\nonumber\\
	&-&\frac12 v_R\, \bar\nu_R^c Y_R \nu_R +{\rm h.c.}
\eea
where $v_R$ is a large mass scale and $v_R Y_R$ gives the Majorana mass matrix. $H$ gets VEV $v/\sqrt{2}$ in its neutral component,
with $v=246$ GeV.
The flavor symmetries imply that
\be
	Y_\nu = \eta'\, Y_e Y_u^\dagger
\label{Ynu}
\ee
We will study the
consequences of this choice for the spectrum of heavy right handed
neutrinos, and the resulting implications for leptogenesis and
low-energy lepton flavor violation.
\bigskip

After integrating out the heavy neutrino, the light
neutrino mass term is
\vskip-0.5cm
\be
	{v^2\over 2v_R}\bar\nu_L\, Y_\nu Y_R^{-1} Y_\nu^T\,
	\nu_L^c + {\rm h.c.}
\ee
Let $\nu_L = L_\nu\nu_m$ denote the relation between the weak eigenstates
$\nu_L$ and the mass eigenstates $\nu_m$.  If $Y_e$ was already
diagonal, then $L_\nu$ would coincide with the PMNS matrix.  However
in a basis where $Y_e$ is not diagonal, this is not the case.
Suppose that the mass term $(v/\sqrt{2})\bar e_L Y_e e_R$ is diagonalized by
taking $e_L\to L_e e_L$, $e_R \to R_e e_R$.  Then the PMNS matrix is
given by
\be
	U_{\sss PMNS} \equiv U =  L_e^\dagger L_\nu
\label{UPMNS}
\ee
where $L_\nu$ diagonalizes the neutrino mass matrix via
\be
	m_\nu = {v^2\over v_R}
	L_\nu^\dagger\, Y_\nu Y_R^{-1} Y_\nu^T\, L_\nu^*
	= {\eta'^2 v^2\over \eta^2 v_R} L_\nu^\dagger\,
	Y_d^\dagger Y_R^{-1} Y_d^*\, L_\nu^*
\label{mnudiag}
\ee

\subsection{Heavy neutrino mass spectra}
\label{spectra}

An interesting feature of the above scenario, also anticipated by ref.\
\cite{Appelquist:2015mga}, is that even if $L_\nu$ is close to the
identity matrix, the factor $L_e^\dagger$ generates large mixing
angles in $U_{\sss PMNS}$.  This means that the Yukawa matrix
$Y_R$ for the sterile neutrino Majorana masses can be very
hierarchical despite the large neutrino mixing angles.

Since $Y_\nu = (\eta'/\eta)Y_d^\dagger$, it is fixed by the down quark
masses, up to an overall normalization factor.  For definiteness, we
choose its largest matrix element to have unit magnitude, anticipating
our application below to leptogenesis where this choice is
advantageous.  It means that the heavy neutrino mass spectra we derive
here represent the maximum sizes consistent with perturbative values
of $Y_\nu$.  Rescaling $Y_\nu$ by a factor of $\lambda < 1$ implies a
reduction in $M_i$ by the factor $\lambda^2$, for fixed values of the
light neutrino masses.

The Majorana matrix $Y_R$ is constrained only by the
experimental values of the neutrino masses and mixing angles.
We can solve eq.\ (\ref{mnudiag}) for $Y_R$, using (\ref{UPMNS}):
\be
	Y_R = {\eta'^2 v^2\over \eta^2 v_R}\,
	(Y_d L_e U)^* \, m_\nu^{-1}\,  (Y_d L_e U)^\dagger
\label{YReq}
\ee
where $m_\nu$ is diagonal and $L_e$ is a unitary transformation such
that $L_e^\dagger\, (Y_e Y_e^\dagger)\, L_e$ is diagonal, using eq.\
(\ref{Ye_pred}) for $Y_e$.  However this determines $L_e$ only up to
multiplication on the right by a diagonal matrix of phases,
$L_e\to L_e\, {\rm diag}(e^{i\beta_1},e^{i\beta_2},e^{i\beta_3})$,
of which two relative phases are physically significant.  In addition,
$U$ can be multiplied on the right by two undetermined Majorana phases.
Therefore $Y_R$ is a function of the down-to-up quark mass ratios,
the CKM parameters, the light neutrino masses, the PMNS parameters,
and four additional phases, as well as the overall scaling factor.

We perform a scan over models where the neutrino mass differences
and mixing angles vary randomly within their allowed ranges, as
specified in appendix \ref{quark_masses}, but the
quark parameters are constrained to be close to values needed to get
the right lepton mass ratios, as described in section \ref{LMR}.
This requires a choice of the lightest neutrino mass $m_{\nu_1}$,
as well as whether the neutrino mass hierarchy is normal or inverted.
We also scan over the four phases mentioned above.  The magnitudes
of the heavy neutrino masses $M_i$ are then fixed, being given by
\be
	M_i = Y_i v_R
\label{Mieq}
\ee
where $Y_i$ are the eigenvalues of $Y_R$ in eq.\ (\ref{YReq}).
 We obtain
distributions of $M_i$ as shown in fig.\ \ref{RHNmasses}, varying
$m_{\nu_1}$ from $10^{-5}$ to $10^{-1}$ eV, and also allowing for
normal or inverted mass hierarchy.  For the majority of models,
there is a clear separation between the mass eigenvalues, and
a hierarchy that becomes more pronounced for smaller values of
$m_{\nu_1}$.  $M_1$ tends to be several orders of magnitude smaller
in the inverted compared to normal hierarchy.

{
The prediction of a hierarchical sterile neutrino spectrum is not
unique to our model.  For example the Altarelli-Feruglio model
\cite{Altarelli:2005yx} which explains tri-bimaximal
mixing as a consequence of the discrete $A_4$ symmetry, predicts
that the heavy neutrino spectrum has the form
$M_i = \{A,\,B-A,\,B+A\}$, while the light neutrino masses go as
$m_i = c/M_i$, so that any hierarchy in the latter is a direct
consequence of hierarchy in the former.  Similarly the
Frampton-Glashow-Yanagida ansatz \cite{Frampton:2002qc}, which has only two sterile
neutrinos, allows for a strong hierarchy between their masses
(although it is not required).  A notable exception is the class
of models based upon minimal lepton flavor violation
\cite{Cirigliano:2005ck} where the flavor symmetry imposes a nearly
degenerate spectrum of heavy neutrinos.}

\begin{figure*}[t]
\includegraphics[width=2\columnwidth]{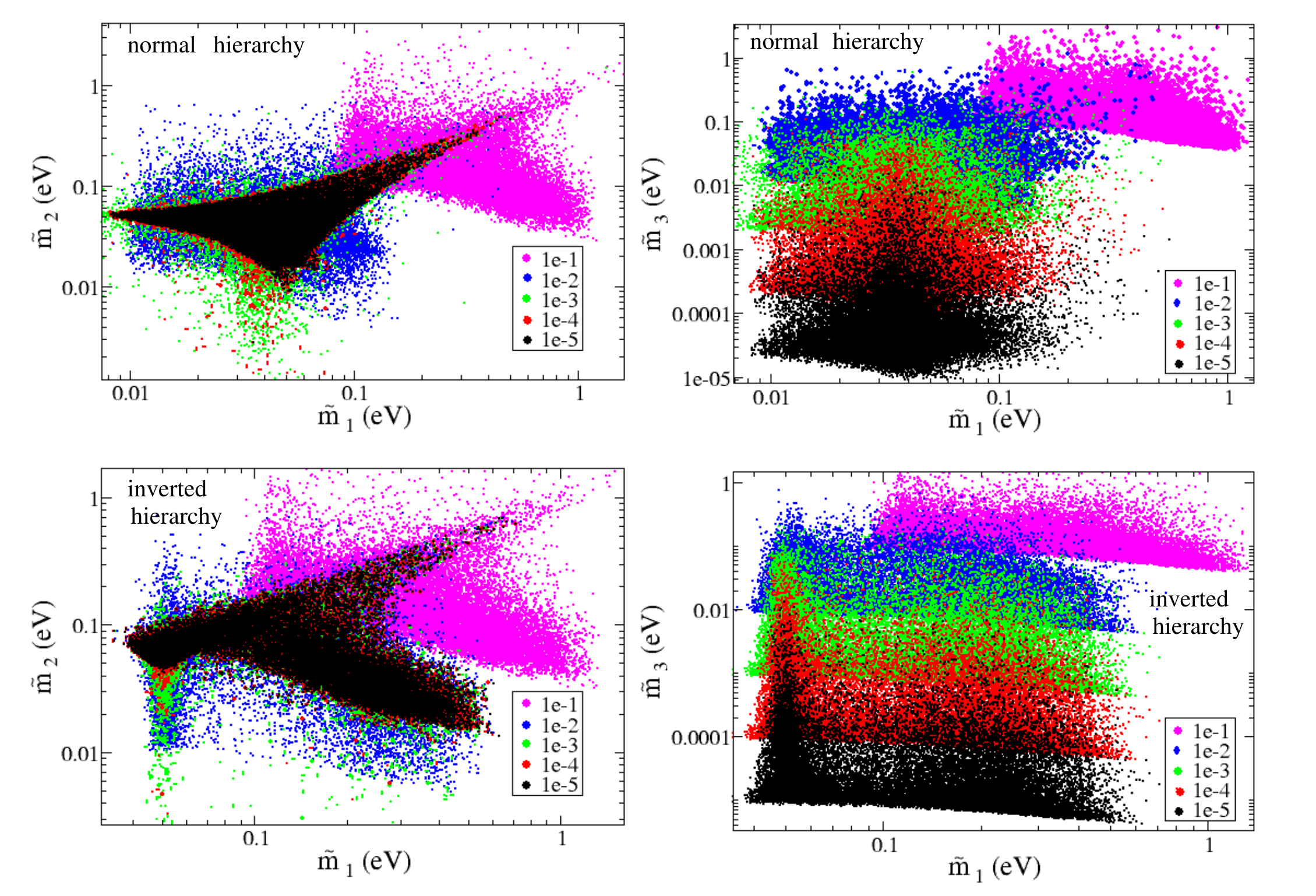}
\caption{Distributions of effective neutrino mass $\tilde m_i$,
eq.\ (\ref{eq:meff}), for different choices of the lightest neutrino
mass $m_{\nu_1} = 10^{-5},\, 10^{-4}\dots,10^{-1}$ eV.  Left (right):
$\tilde m_2$ ($\tilde m_3$) versus $\tilde m_1$.  Top (bottom)
Normal (inverted) mass hierarchy.
Smaller values of $m_{\nu_1}$ correspond to smaller values of
$\tilde m_3$.
}
\label{meff_fig}
\end{figure*}

\begin{figure*}[t]
\includegraphics[width=2\columnwidth]{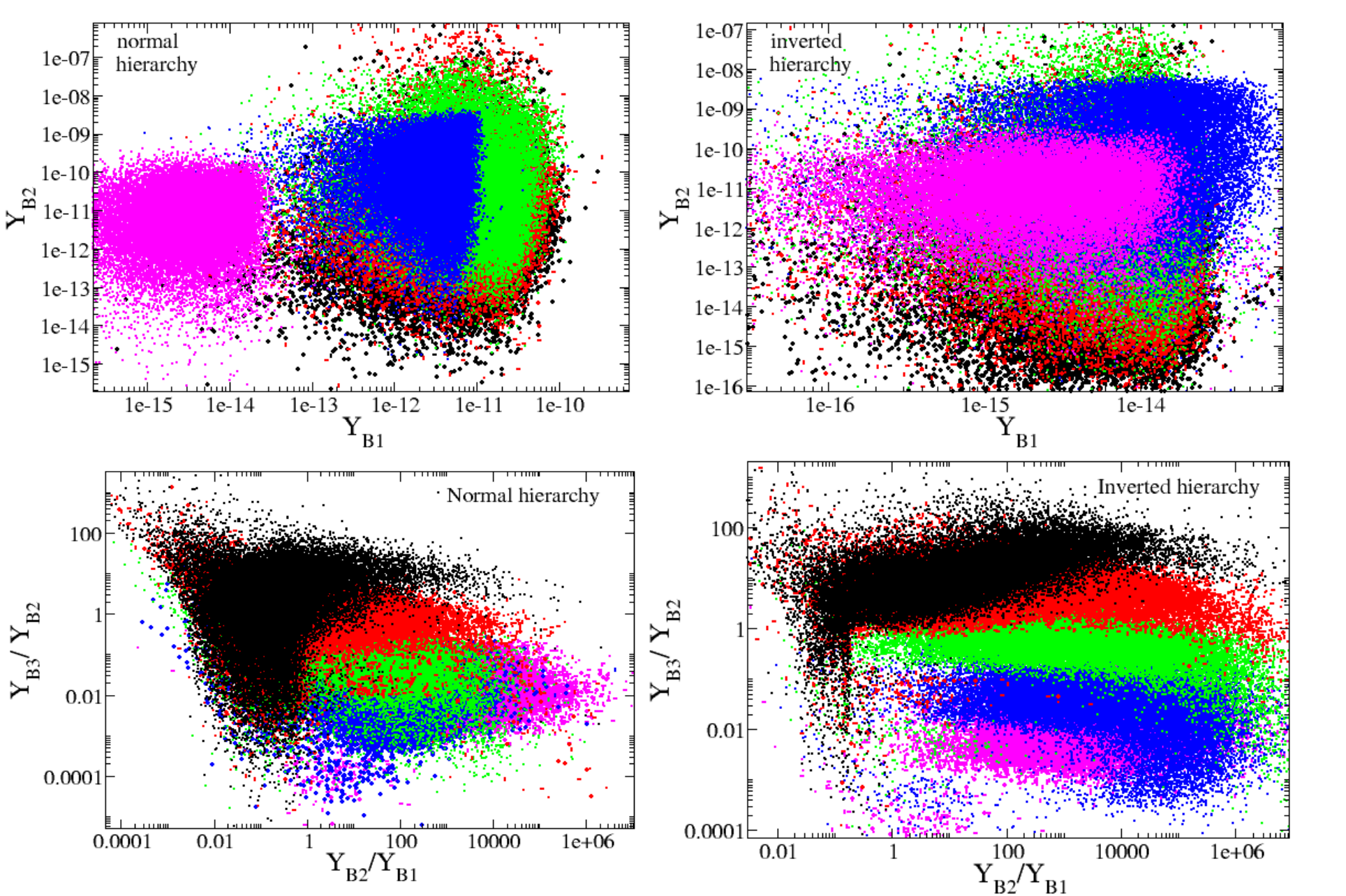}
\caption{Top row: distributions of baryon asymmetries $Y_{B2}$ versus
$Y_{B1}$ produced by $N_2$
versus $N_1$ decays for normal hierarchy (left) and inverted
hierarchy (right).  Colors correspond to light neutrino mass
values as in previous figures.  $Y_{B3}$ versus $Y_{B1}$
(not shown) is similar.  Bottom row: distributions of abundance
ratios $Y_{B3}/Y_{B2}$ versus $Y_{B2}/Y_{B1}$, before accounting
for washout by $N_1$ and $N_2$ interactions.
}
\label{YB2_fig}
\end{figure*}

\section{Leptogenesis}

{{
As an application of CFV extended to include neutrinos, we explore the
consequences for thermal leptogenesis from taking the ansatz (\ref{Ynu}) for
the neutrino Dirac Yukawa matrix.  Thermal leptogenesis assumes that
the heavy neutrino whose decays produce the asymmetry is initially
not present in the thermal bath, but only gets generated by its Yukawa
interactions \cite{Buchmuller:1996pa,Plumacher:1996kc}.   This
assumption allows for definite predictions, which would otherwise be
ambiguous due to dependence upon the initial conditions.

It is important to notice that the CFV ansatz only fixes
the ratios of right-handed neutrino masses $M_i$ but not their overall
scale. On the other hand the asymmetry produced by leptogenesis can
depend strongly on the actual scale of $M_i$. To gain a qualitative
understanding of possible  correlations between the lepton asymmetry
and the structure of couplings predicted by CFV, free from
uncertainties from the undetermined overall scale, we first focus on
the simplest scenario of thermal leptogenesis (for recent reviews see
\cite{Chen:2007fv,Fong:2013wr}). Namely, we assume the reheating
temperature $T_\textrm{rh}$ is much higher than $M_i$,  and $N_i$ decays in
the ``single lepton flavor regime.''  This means that the linear
combination of flavors produced by the decay of heavy neutrino $N_i$
does not decohere due to Yukawa interactions.  In other words, it
is assumed that the temperature is sufficiently high for these
interactions to not yet have come into equilibrium.

For the hierarchical $N_i$ mass spectrum, it is often assumed that
interactions mediated by the lightest neutrino $N_1$ wash out any
lepton asymmetries produced by decay of $N_{2,3}$, in which case the
one produced by $N_1$ decay is relevant.  However under certain
conditions, $N_1$ decays can only destroy a particular linear
combination of lepton number, that only partially overlaps with the
combinations created in $N_{2,3}$ decays
{{\cite{Vives:2005ra}}}\cite{Engelhard:2006yg}.

For an initial estimate, we will calculate the asymmetries that can be
produced by any of the three decays in the simplest scenario in
section \ref{sec:single}, ignoring washout effects mediated by the lighter
neutrinos. We then comment on the flavor effects in
section \ref{sec:flavor} when $N_i$ decays at lower temperature. We
discuss the pure $N_1$ contribution in section \ref{sec:N1lepto}. The case
that $N_{2,3}$ decay dominates is discussed in
section \ref{sec:N2N3lepto}, with results summarized in
section \ref{sec:results}.  In section \ref{sec:comp} we compare our
results to those of other scenarios for leptogenesis.
In this work we focus on the qualitative
features and are not concerned with $\mathcal{O}(1)$ uncertainties.}}

\subsection{Decoupled asymmetries}
\label{sec:single}

Essential quantities for estimating the lepton asymmetry from heavy
neutrino $N_i$ decays are the CP asymmetries $\epsilon_i$.  These are expressed in the basis where
$Y_R$ is diagonal (with positive real entries), $\nu_R = R_\nu N_i$, implying
$R_\nu^\dagger(Y_R^\dagger Y_R) R_\nu$ is diagonal, and its
eigenvalues $Y_i^2$ determine the heavy neutrino masses through
eq.\ (\ref{Mieq}).  In the special basis assumed for eq.\
(\ref{Ye_pred}), $Y_d$ and $Y_\nu$ are both diagonal.  In the
basis where $Y_R$ is diagonal, the neutrino Yukawa matrix
is $h = R_\nu^\dagger
Y_\nu^\dagger$.  Then the CP-asymmetry for decay of heavy neutrino
$N_i$ is
\be
	\epsilon_i = {1\over 8\pi (h h^\dagger)_{ii}}
	\sum_{j\neq i} {\rm Im}(h h^\dagger)^2_{ji} \,
	g(M_j^2/M_i^2)
\label{epseq}
\ee
where $g(x) = \sqrt{x}[1/(1-x) + 1 - (1+x)\ln(1+1/x)]$.  It is maximized when
$h$ is as large as possible, hence for our initial estimates of the
maximum possible asymmetries we scale $Y_\nu$ so that the
largest element of $h_{ij}$ is unity (as in section \ref{spectra}),
still consistent with a perturbative analysis.

The baryon asymmetry is conveniently expressed as $Y_B$,
the baryon-to-entropy ratio. Big bang nucleosynthesis and the cosmic microwave
background give consistent determinations,
$Y_{B,\textrm{obs}}\cong 8\times 10^{-11}$ \cite{BBN}\cite{CMB}.
The initially-produced asymmetry can be parametrized as
\cite{Davidson:2008bu}
\be\label{eq:etaB}
	{{Y_B}} = \sum_i Y_{B,i}
	\ \sim \ 0.4\,{c_{\rm sph}\over  g_*}\,\sum_i
	\kappa_i\,
	\epsilon_i
\ee
where $c_{\rm sph}$ depends upon how sphalerons redistribute the lepton
asymmetry ($c_{\rm sph} = 28/79$ in the standard model),
$g_*=106.75$ degrees of freedom in the SM plasma, and $\kappa_i$ is
an efficiency factor taking account of washout of the produced lepton
asymmetry, due to rescatterings mediated by the decaying neutrino
$N_i$\footnote{{{The estimate in eq.(12) assumes thermal leptogenesis ends
before the sphaleron processes become active and $Y_{B-L}$ is
transferred to $Y_B$ by sphalerons later. In the case that
$10^9\lesssim M_i\lesssim 10^{12}$\,GeV, electroweak sphalerons are in
equilibrium with other processes, which may cancel a (small) part of
the lepton asymmetry that contributes to
$Y_{B-L}$~\cite{Engelhard:2006yg}. We ignore this $\mathcal{O}(1)$
uncertainty in the following discussion.}}}.

The efficiency factor $\kappa_i$ is a function of the ratio of the
$N_i$ decay rate to the Hubble rate,
\bea
\label{Kieq}
K_i&=&\frac{\Gamma_{N_i}}{H(M_i)}
	= {M_{pl}\over (1.66)(8\pi\sqrt{g_*})}{(h h^\dagger)_{ii}
	\over M_i}
\eea
where $\Gamma_{N_i}=\frac{1}{8\pi}(hh^\dag)_{ii}M_i$,
$H=1.66\sqrt{g_*}\,{T^2}/{M_{pl}}$, and $M_{pl}=1.22\times
10^{19}\,$GeV is the Planck mass.  The rates are proportional to
the effective light neutrino masses $\tilde{m}_i$, defined as
\begin{eqnarray}
\label{eq:meff}
 \tilde{m}_i=\frac{(hh^\dag)_{ii}v^2}{M_i}
\end{eqnarray}
The respective conditions $K_i\ll1$ and
$K_i\gg 1$ for weak and strong washout
correspond to $\tilde m_i \ll m_*$ and $\tilde m_i \gg m_*$,
with $m_*\simeq 10^{-3}\,$eV.

In general, we have the freedom to scale $M_i$ and $Y_\nu^2$
(or $h h^\dagger$) by a
common factor while keeping the light neutrino mass spectrum
fixed.  The washout factors $K_i$ are invariant under this scaling,
while the asymmetries $\epsilon_i$ have a linear dependence.
Hence we have chosen $M_i$ and $h h^\dagger$ to be as large
as is consistent with a perturbative treatment, ${\rm max}|h_{ij}|=1$,
for our estimate of the initially produced asymmetries.
This freedom can be used to dial down the asymmetry in cases where it
may be larger than the observed one.
For {{our}} initial estimates, we follow ref.\ \cite{Buchmuller:2004nz} in estimating the washout
factor as
\be\label{eq:kappai}
	\kappa_i \cong {\rm min}\left( 2\times 10^{-2}
	\left({0.01{\rm\, eV}\over
	\tilde m_i}\right)^{1.1},\ 1\right)
\ee
In the weak-washout regime, the produced asymmetries are sensitive
to the initial values of the heavy neutrino abundances.
The approximation $\kappa_i=1$ in this case assumes a thermal
distribution.

As before, we scan over quark parameters consistent with the
lepton mass ratios, and over neutrino masses and mixing angles
as given in appendix \ref{quark_masses}.  Distributions of
the effective neutrino masses $\tilde m_i$ are shown in fig.\
\ref{meff_fig}.  For almost all cases, $\tilde m_{1,2} \sim 0.01-1{\rm\, eV}
> m_*$, corresponding to some washout of the $N_{1,2}$ asymmetries,
while $\tilde m_3$ can be in the weak washout regime, if $m_{\nu_1}
\lesssim 10^{-4}$ eV.  In fig.\ \ref{YB2_fig} we show the
contributions to the baryon asymmetry $Y_{B2}$ versus $Y_{B1}$ from
the same decays. (The results for $N_3$ decays look similar to those
from $N_2$.)   For the normal hierarchy, $Y_{B1}$ can be marginally
big enough to account for observations, but it falls short in the
inverted hierarchy.  On the other hand $Y_{B2}$ tends to give the
dominant contribution (assuming it does not get washed out by
subsequent $N_1$ scatterings), motivating us to give more careful
consideration to this scenario below.

We have also performed scans using values of the neutrino mixing angles
at the GUT scale.  The exact values depend upon the structure of the
lepton and neutrino Yukawa couplings, and the heavy neutrino masses,
which give thresholds in the renormalization group equations
\cite{Antusch:2005gp}.  It is beyond the scope of this work to carry
out the RGE evolution for each model in our scans; instead we adopted
typical GUT scale inputs $\sin^2\theta_{12} = \sin^2\theta_{23} =
0.5$, $\sin^2\theta_{13} < 0.01$ (while the running of the light
neutrino masses is small enough to neglect)
found in a survey of neutrino
models \cite{Gupta:2014lwa}.  The results do not differ markedly from
those presented above from the low-energy neutrino parameters.

\subsection{Flavor effects}
\label{sec:flavor}

The simple description given above needs to be modified in a more
quantitative treatment taking account of flavor effects. We so far
assumed that  $N_i$ decays in the single lepton flavor regime, where
the lepton state produced by $N_i$ decay propagates coherently.  This
is true at sufficiently high temperatures such that interactions
involving the charged lepton Yukwawa couplings are out of equilibrium.
These come into equilibrium as the temperature decreases:
the equilibration temperatures for the $e,\mu,\tau$ couplings are given
by $T_e\simeq 4\times 10^{4}\,$GeV, $T_\mu\simeq 2\times 10^9\,$GeV,
$T_\tau\simeq 5\times 10^{11}\,$GeV respectively \cite{Fong:2013wr}.

Therefore for $T\lesssim 10^9$\,GeV, the flavor basis $\{\ell_e,
\ell_\mu, \ell_\tau\}$ gets fully resolved by scattering processes
involving the Higgs.  On the other hand, in the temperature window
$10^9\lesssim T\lesssim 10^{12}\,$GeV, the lepton state produced from
$N_i$ decay gets projected onto the $\ell_\tau$ direction and some
linear combination of light flavors $\ell_{e+\mu}$.
Since $N_j$ (with $j\le i$) can only completely wash out the flavor direction
to which it couples, some parts of these asymmetries are partially
protected from washout.

When the Boltzmann equations are modified to take account of  this
effect, the net washout is reduced relative to the naive  treatment,
and the  final asymmetry is typically enhanced. When only the $y_\tau$
Yukawa coupling is important, the symmetry is generally enhanced by a factor of
2 relative to the single flavor case \cite{Fong:2013wr}. In parts of
parameter space, it is possible to get an asymmetry purely from flavor
effects even when $\epsilon_i=0$ \cite{Nardi:2006fx}. For our more
quantitative  numerical study of $N_i$ leptogenesis for CFV, rather
than solving the exact Boltzmann equations for the case
$M_i<10^{12}\,$GeV, we roughly estimate the tau flavor effect by multiplying
the asymmetry in (\ref{eq:etaB}) by a factor of 2.

\subsection{$N_1$ leptogenesis}
\label{sec:N1lepto}

Since our heavy neutrino mass spectrum is very hierarchical, it is
plausible that the reheating temperature $T_{\textrm{rh}}$ is in between $M_1$ and
$M_{2,3}$, so that $N_1$ has a thermal distribution but $N_{2,3}$
are highly suppressed in their abundances.
However in this case the likelihood of sufficient
baryogenesis from $N_1$ decays alone is quite small, as can be seen from
fig.\ \ref{YB2_fig}.  Only with the normal mass hierarchy is it possible.
For $m_{\nu_1} = 10^{-5}\,$eV, 1 in 1000 random
models have $Y_{B1}$ as large as the observed value, and this fraction
remains constant with increasing $m_{\nu_1}$, until  $m_{\nu_1} \sim
0.01\,$eV when it drops to zero.

The heavy neutrino masses $M_1$ cluster around their maximum values
in the interval $(0.5-2)\times 10^{11}$ GeV for these models,
which also have relatively large values of $|\epsilon_1|$, compared
to the average.  The preference for large $M_1$ and the scarcity of
viable models can be understood in terms of the strong washout effect
from the $N_1$ neutrinos.  Higher efficiency requires smaller values
of $\tilde m_1 \sim M_1^{-1}$.

From fig.\ \ref{RHNmasses} it is clear that all models have
$M_1<10^{12}\,$GeV. Hence the $N_1$-generated asymmetry  estimated in
eq.\ (\ref{eq:etaB}) may get enhanced by a factor of $\mathcal{O}(1)$  as
mentioned in section \ref{sec:flavor}, due to the $\tau$ flavor effect. This small
enhancement does not change the qualitative  unlikelihood of successful
$N_1$ baryogenesis in this model.

\subsection{$N_{2,3}$ leptogenesis}
\label{sec:N2N3lepto}

On the other hand, a large fraction of models in our scans can produce
a sufficient baryon asymmetry through the decays of
$N_2$
or $N_3$ neutrinos. We will first describe that coming from
$N_2$, then explain its generalization to $N_3$.

\vspace{1.5em}

As first pointed out by ref.\ {{\cite{Vives:2005ra}}}\cite{Engelhard:2006yg},
a large fraction of the
asymmetry produced by $N_2$ can survive the effects of
$N_1$-mediated scatterings even if the latter have not decoupled,
but are in fact fast enough to
induce decoherence of the $N_2$-produced lepton state.
In this case, it is only a projection of the original state that is
washed out by $N_1$ interactions.

The conditions for strong $N_1$  washout (decoherence) are
\begin{eqnarray}\label{eq:washoutCond}
M_2\gg M_1,\quad
\tilde{m}_1\gg m_*,\quad
\frac{M_1}{M_2}<\frac{m_*}{\tilde{m}_2}
\end{eqnarray}
where we choose $M_2>10\,M_1$ and $\tilde{m}_1>10\,m_*$ in practice.
The
third condition ensures that $N_2$-mediated interactions
play a negligible role at $T\sim M_1$ compared to the $N_1$ decoherence
effect. A further condition for protecting a direction in flavor
space from
$N_1$ washout is that $M_1>10^9\,$GeV, so that it does not decohere
fully into its $\{\ell_e,
\ell_\mu, \ell_\tau\}$ flavor components.
The lepton doublet
produced in decays of $N_i$ is initially in the flavor superposition
\be
|\ell_i\rangle=(hh^\dag)^{-1/2}_{ii}\sum_\alpha h_{i\alpha}|\ell_\alpha\rangle
\label{elli} \equiv \sum_\alpha c_{\alpha i} |\ell_\alpha\rangle
\ee
where $i=1,2,3$,  $\alpha = e,\mu,\tau$ and $c_{\alpha i} = \langle
\ell_\alpha|\ell_i\rangle$.  In the following we use roman index $i$
for the heavy neutrino mass eigenbasis, and greek index $\alpha$ for
the flavor basis.

As mentioned before, the lepton flavor effect from $y_\tau$ Yukawa
interactions becomes relevant when the temperature is below
$10^{12}\,$GeV. This is always true for $N_1$ decays, since
our preliminary scan (fig.\,\ref{RHNmasses}) shows that
$M_1 < 10^{12}\,$GeV in all cases.  Thus only the projection
of $|\ell_i\rangle$ orthogonal to $|\ell_\tau\rangle$ remains coherent.
Of this, the part that is orthogonal to $|\ell_1\rangle$ is untouched
by $N_1$ washouts, while the parallel component is reduced by the
efficiency factor $\kappa_1$.

To quantify the resulting baryon asymmetry, we define an orthonormal
basis $\{\ell_\tau,\, \ell_0,\, \ell'_1\}$,
\begin{eqnarray}\label{eq:basis2}
   |\ell_0\rangle&=& N_1'
   \left(c_{\mu1}^*|\ell_e\rangle-c_{e1}^*|\ell_\mu\rangle\right)\nonumber\\
   |\ell'_1\rangle&=&N_1'
   \left(c_{e1}|\ell_e\rangle+c_{\mu1}|\ell_\mu\rangle\right)
\end{eqnarray}
where $N_1' = (|c_{e1}|^2+|c_{\mu1}|^2)^{-1/2}$,
so that $|\ell_2\rangle$ decomposes as
\begin{eqnarray}
\label{eq:l2p2}
   |\ell_2\rangle=c_{\tau 2}|\ell_\tau\rangle
   +c_{0 2}|\ell_0\rangle
   +c'_{12}|\ell'_1\rangle
\end{eqnarray}
with {{$c_{02}=\langle\ell_0|\ell_2\rangle$, $c'_{12}=\langle\ell'_1|\ell_2\rangle$.}}
Then the corrected asymmetry is
\be
\label{eq:YB2eff}
 Y_{B2}\cong
   Y_{B2,0}\left(|c'_{12}|^2\kappa_{1}+|c_{02}|^2\right)
\ee
where $Y_{B2,0}$ is the naive estimate given in eq.\ (\ref{eq:etaB}),
if $M_2>10^{12}\,$GeV.  If $M_2<10^{12}\,$GeV, we include an
extra factor of 2 to estimate the
$y_\tau$ Yukawa effect as discussed above.
For this quantitative study we use the more exact approximations
for the efficiency factors from ref.\ \cite{Buchmuller:2004nz},
given in appendix \ref{eff_fact}.


\vspace{1.5em}

An analogous procedure can be carried out to include the contribution
from $N_3$ decays to the surviving asymmetry.
{According to fig.\ \ref{RHNmasses}, if $M_1$ is no smaller than
$10^9\,$GeV, we always have $M_3>10^{12}$\,GeV. Thus
 the initial $N_3$-generated asymmetry $Y_{B3,0}$
is given by eq.\  (\ref{eq:etaB}).
If $N_2$ interactions are fast, they will wash out the
part of  $|\ell_2\rangle$ that is parallel to
$|\ell_3\rangle$.
The analogous  condition to eq.\ (\ref{eq:washoutCond}) is
\begin{eqnarray}\label{eq:washoutCond2}
M_3\gg M_2,\quad
\tilde{m}_2 \gg m_*,\quad
\frac{M_2}{M_3}<\frac{m_*}{\tilde{m}_3}
\end{eqnarray}
Again, we demand that $M_3>10\,M_2$ and $\tilde{m}_2>10\,m_*$,
similarly to the conditions (\ref{eq:washoutCond}).
Depending upon whether $M_2>10^{12}\,$GeV, the appropriate
expansions are
\be\label{eq:basis30}
	|\ell_3\rangle = \left\{
	{\ \ \ \ \ \  \qquad c_{23}|\ell_2\rangle + c'_{03}|\ell_0'\rangle,\quad
	M_2 > 10^{12}{\rm\ GeV}\atop
	c_{\tau 3}|\ell_\tau\rangle + c'_{23}|\ell'_2\rangle
	+ c''_{03}|\ell_0''\rangle,\quad
	M_2 < 10^{12}{\rm\ GeV} }\right.
\ee
where $c'_{03} = (1-|c_{23}|^2)^{1/2}$ (we are free to define the
phase of $\ell'_0$ such that $c'_{03}$ is real and positive) and
{{\begin{eqnarray}
\label{eq:basis3}
   |\ell_0'\rangle&=&\left(|\ell_3\rangle-c_{23}|\ell_2\rangle\right)/c_{03}^{\prime}\nonumber\\
   |\ell_0''\rangle&=&N_2'
   \left(c_{\mu2}^*|\ell_e\rangle-c_{e2}^*|\ell_\mu\rangle\right)\nonumber\\
   |\ell'_2\rangle&=&N_2'
   \left(c_{e2}|\ell_e\rangle+c_{\mu2}|\ell_\mu\rangle\right)
\end{eqnarray}}}
with $N_2' = (|c_{e2}|^2+|c_{\mu2}|^2)^{-1/2}$,
{{and $c''_{03}=\langle\ell''_0|\ell_3\rangle$, $c'_{23}=\langle\ell'_2|\ell_3\rangle$.}}

The components $|\ell_2\rangle$, $|\ell'_2\rangle$ get suppressed
by $\kappa_2$, while $|\ell'_0\rangle$, $|\ell''_0\rangle$ are
unaffected by $N_2$ washouts.
{If the $N_1$ strong washout condition (\ref{eq:washoutCond}) is
satisfied,}}
all of these basis vectors must then be
expanded in terms of $|\ell_0\rangle$ and $|\ell_1'\rangle$ to
determine the effect of $N_1$ washouts.  This was already done for
$|\ell_2\rangle$ in eq.\ (\ref{eq:l2p2}).  For the rest,
\bea
\label{eq:others}
   |\ell'_0\rangle &=& c'_{\tau 0}|\ell_\tau\rangle
   +c'_{0 0}|\ell_0\rangle  +c''_{10}|\ell'_1\rangle \nonumber\\
   |\ell''_0\rangle &=& c''_{\tau 0}|\ell_\tau\rangle
   +c''_{0 0}|\ell_0\rangle  +c'''_{10}|\ell'_1\rangle \nonumber\\
   |\ell'_2\rangle &=& c'_{\tau 2}|\ell_\tau\rangle
   +c'_{0 2}|\ell_0\rangle  +c''_{12}|\ell'_1\rangle
\eea
{{where $c'_{00}=\langle\ell_0|\ell'_0\rangle$,
$c''_{10}=\langle\ell'_1|\ell'_0\rangle$,
$c''_{00}=\langle\ell_0|\ell''_0\rangle$,
$c'''_{10}=\langle\ell'_1|\ell''_0\rangle$,
$c'_{02}=\langle\ell_0|\ell'_2\rangle$,
$c''_{12}=\langle\ell'_1|\ell'_2\rangle$ with the explicit
expressions for the state vectors
in (\ref{eq:basis2}, \ref{eq:basis3}, \ref{eq:others}).}} The naive
contribution to $Y_{B3}$ gets reduced analogously to
(\ref{eq:YB2eff}) as $Y_{B3}\cong P_3 Y_{B3,0}$ where
\bea\label{eq:P3L}
	P_3 &=&  \ \ \ \ \kappa_2\, |c_{23}|^2 \left( |c_{02}|^2 +
	\kappa_1 |c'_{12}|^2\right)\nonumber\\
	&+& (1-|c_{23}|^2)\left(|c_{00}'|^2 + \kappa_1|c''_{10}|^2
	\right)
\eea
for $M_2 > 10^{12}\,$GeV, and
\bea\label{eq:P3S}
	P_3 &=&  \kappa_2\, |c'_{23}|^2 \left( |c'_{02}|^2 +
	\kappa_1 |c''_{12}|^2\right)\nonumber\\
	&+& \ \ \ \ |c''_{03}|^2\left(|c_{00}''|^2 + \kappa_1|c'''_{10}|^2
	\right)
\eea
for $M_2 < 10^{12}\,$GeV.

\subsection{Results for $N_{2,3}$ leptogenesis}
\label{sec:results}

Here we present results for successful $N_{2,3}$ leptogenesis from CFV
model samples based on the formalism discussed in previous  subsection.
First let us estimate how readily the two strong washout conditions
(\ref{eq:washoutCond}) and (\ref{eq:washoutCond2}) can be satisfied.
These conditions are independent of an overall rescaling of neutrino
Yukawa couplings and heavy masses (in which the light neutrino masses
remain fixed), so we can analyze them prior to doing such a rescaling,
which we will use to renormalize any too-large baryon asymmetry down to
the observed value. As shown by the parameter scan in figs.\
\ref{RHNmasses} and \ref{meff_fig}, we find $\tilde{m}_i\gg m_*,\,
M_{i+1}\gg M_i$ with $i=1,2$ hold almost always.  Hence the fraction of
models meeting the strong washout requirements is mainly determined by
the last condition in each case. For the five representative values of
$m_{\nu_1}$ considered, we find that more than $90\%$ of samples pass
conditions (\ref{eq:washoutCond2}),  while roughly $70\%$
satisfy (\ref{eq:washoutCond}). The fraction with both conditions
satisfied is $\sim 60-70\%$.

We select samples that successfully generate observed baryon asymmetry
by $N_{2,3}$ decay in following way. Assuming $M_2>10^{12}\,$GeV
initially, we first check the validity of the $N_2$ strong washout
condition (\ref{eq:washoutCond2}). If it is satisfied, we make the
$|\ell_3\rangle$ projection as in (\ref{eq:basis30}), otherwise $N_3$
decay does not contribute in the final asymmetry. Next we check $N_1$
strong washout condition (\ref{eq:washoutCond}). If it is satisfied, we
have $Y_B=Y_{B2}+Y_{B3}$ with projections (\ref{eq:YB2eff},
\ref{eq:P3L}), otherwise we reject the sample model.

In any case, we require $r = Y_B/Y_{B,\textrm{obs}} \ge 1$ since we have found
the largest possible asymmetry, corresponding to a prescribed value of
the maxiumum allowed neutrino Yukawa coupling $|h_\nu^{ij}|$.  Two
choices are considered, $\max |h_\nu^{ij}| = 1$ and
$\sqrt{4\pi}$, the latter being the largest allowed by
perturbative unitarity.  We then rescale $y_\nu$ by $1/\sqrt{r}$ and
$M_i$ by $1/r$ to bring $Y_B$ into agreement with the observed value.
The above consistency requirements with respect to the heavy
neutrino masses must be rechecked following this rescaling before
declaring the sample model to be successful.

\begin{figure}[t]
\includegraphics[width=\columnwidth]{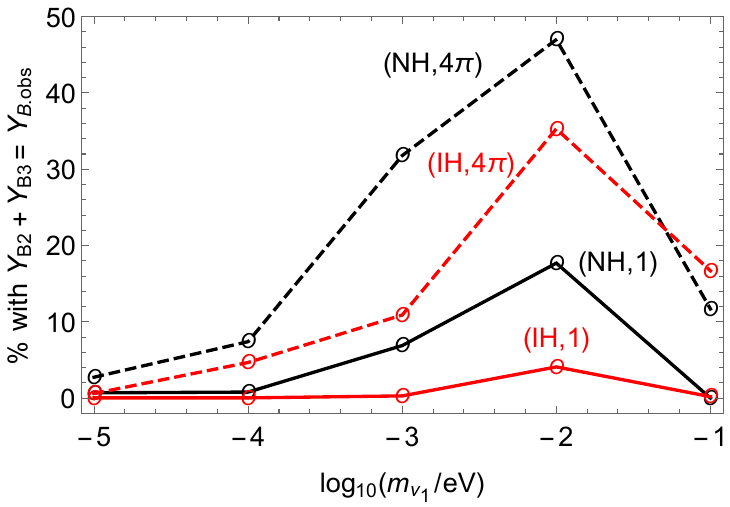}
\caption{Percentage of random scans yielding successful $N_2, N_3$ baryogenesis
as a function of lightest neutrino mass for normal (black) and inverted (red) hierarchy. For each case, we consider two choices of max $|h_\nu^{ij}|^2=1$ (solid) and $4\pi$ (dashed).
}
\label{goodYB23fig}
\end{figure}

We present the fraction of successful $N_{2,3}$ baryogenesis
models in fig.\
\ref{goodYB23fig}, as a function of the lightest neutrino mass,
and for the two choices $1,\sqrt{4\pi}$ of ${\rm max}|h_\nu^{ij}|$.
Generally, the fraction increases with $m_{\nu_1}$, but drops abruptly
above $\sim 0.01$ eV. The negligible fraction at $0.1\,$eV is
correlated with the stronger $N_1$ washout effect indicated in
fig.\ \ref{meff_fig}.  The yield is consistently
greater in the case of normal mass hierarchy, where the models tend
to have larger values of $M_1$ than for the inverted hierarchy.
$N_2$ decay almost always gives the dominant contribution
to the final asymmetry. Only when
$m_{\nu_1}\lesssim 10^{-5}\,$eV, the weak $N_3$ washout effect becomes
important and $N_3$ decay contributes significantly in the selected
samples\footnote{When $M_3$ is quite large, the gauge
and top interactions might not be in equilibrium. Ref. \cite{Davidson:2008bu} (see Table 1) shows the corrections to the yield of leptogenesis assuming certain spectator processes are in equilibrium. Generally they induce $\mathcal{O}(1)$ uncertainty for the estimation based on
eq.\ (\ref{eq:etaB}) and (\ref{eq:kappai}).}. In summary, the CFV extension to neutrinos provides a
framework for a very hierarchical right-handed neutrino mass spectrum,
where the observed baryon asymmetry mainly comes from $N_{2,3}$ decay.
The fraction of successful models, taken from random samples consistent with
the charged lepton mass spectrum, can be significant, $20-50\%$,
especially if $m_{\nu_1} \sim 10^{-2}\,$eV.


\subsection{Comparison to other frameworks}
\label{sec:comp}

To conclude our study of leptogenesis, we compare our results with
previous literature based on other ansatzes for the neutrino Yukawa
couplings. In the Altarelli-Feruglio model, it was found that the
CP-asymmetry vanishes at leading order as the consequence of the $A_4$
symmetry for tribimaximal mixing~\cite{Jenkins:2008rb}. Hence the
lepton asymmetry can only be generated by subleading
corrections. An extension of the Altarelli-Feruglio model was
recently studied in ref.\ \cite{AF}, where it was shown that
with the right-handed neutrino decaying in
the single lepton flavor regime, a sufficient lepton
asymmetry can be generated by next-to-leading order terms. However
washout effects mediated by the lighter neutrino were not considered,
which are crucial in our estimates for the CFV framework.

For the Frampton-Glashow-Yanagida model, a systematic study was
recently done in ref.\ \cite{FGY}, where it was shown that with
renormalization group running, the normal neutrino mass hierarchy is
disfavored. In the inverse hierarchy case, successful leptogenesis can
be realized with $M_1\sim 10^{13}$\,GeV. The naturalness of the
125\,GeV Higgs boson prefers a much lower mass of heavy
neutrino~\cite{Clarke:2015gwa}. Taking this into account, the authors
of \cite{FGY} studied resonant leptogenesis with a nearly-degenerate
mass spectrum, finding that sufficient baryon asymmetry
can be generated if
$M_1 < 4 \times 10^7$\,GeV.

A class of minimal seesaw models involving  two right-handed neutrinos
and Yukawa matrices with one texture zero was recently studied in
ref.\ \cite{King} with respect to its predictions for  leptogenesis.
There is only one important phase in this scenario, that controls both
leptogenesis and low-energy CP violation; the predicted CP phase is
still consistent with current experimental data.
An example of an SU(5) SUSY
GUT model was studied with the lepton asymmetry coming mainly from the
lightest right-handed neutrino decay. This is in contrast to the
dominance of $N_2$ leptogenesis in our case.

For realizations of leptogenesis based upon minimal lepton flavor
violation \cite{Cirigliano:2005ck}, the heavy right-handed neutrino
masses are degenerate at tree level, with mass differences only
generated radiatively by loop corrections. The lepton asymmetry is then derived
from a different mechanism, resonant leptogenesis, which is unlikely
to be realized in CFV. Successful leptogenesis was obtained with the
right-handed neutrino mass scale above $10^{12}$\,GeV. This implies
sizable neutrino Yukawa couplings and the possibility to probe the
additional new physics scale up to 100\,TeV via lepton flavor
violation in low energy
experiment~\cite{Cirigliano:2006nu}\cite{Pilaftsis:2015bja}.

\section{Lepton flavor violation}

By the assumed symmetries of constrained flavor violation, there
can be exotic dimension-6 operators that are suppressed only by
heavy mass scales and no Yukawa couplings,
\be
	{1\over\Lambda_1^2}|\bar e^{i}_{\sss R} \gamma^\mu u^i_{\sss R}|^2,\quad
	{1\over\Lambda_2^2}|\bar d_{\sss R}^i L_{\sss L}^i|^2
\label{leptoquark}
\ee
where color indices are implicit. These have the structure of
leptoquark-induced interactions, which
have been previously studied in the context of MFV interactions
in refs.\
\cite{Arnold:2009ay,Davidson:2010uu}. Both operators lead to lepton-flavor violating
and  quark $\Delta F=1$ decays, $D\to e\mu$ from the first operator
and $K\to e\mu$, $B_d\to e\tau$ and $B_s\to \mu\tau$
from the second one.  Ref.\ \cite{Appelquist:2015mga} chose not to
consider these interactions; we do so here.

Of the rare leptonic decays, the most highly constrained
is $K^0_L\to \mu e$ with branching ratio $< 5\times 10^{-12}$.
Corresponding constraints on the new physics scale have been worked
out in ref.\ \cite{Davidson:1993qk}.  Updating their result with the
current experimental limit reported in PDG \cite{PDG}, we find
\be
	\Lambda_2 > 260{\rm\, TeV}
\ee
which is comparable to the bound derived from MFV leptoquarks that
have Yukawa structure \cite{Davidson:2010uu}.
The corresponding constraint from $D^0\to e\mu$ with branching ratio
$< 2.6\times 10^{-7}$ is much weaker, since it is helicity
suppressed by $m_\mu/m_D$, and the constraint on the branching ratio
is also weaker:
\be
	\Lambda_1 > 1.7{\rm\, TeV}
\ee

In addition to the purely leptonic channels, there are semileptonic
decays such as $K^0_L\to \pi\mu e$ that are also strongly constrained
by experiment.  However ref.\ \cite{Davidson:1993qk} finds these
generally less constraining on the scales $\Lambda_i$ than the purely
leptonic ones.

\section{Conclusions}

Evidence for a simple constraint $Y_d\sim Y_u Y_e^\dagger$ between the
fermion Yukawa matrices of the standard model would be very
interesting, for gaining insights into the origin of flavor. The
biggest challenge for this hypothesis is that charged lepton masses
are known extremely well, and do not agree with the  naive predictions
coming from this relation.  We have shown that if the prediction
actually applies at a high scale such as the GUT scale, and if the
up-to-down quark mass ratio is somewhat larger at this scale than at
low energies, the problem with lepton masses can be overcome.  It is
conceivable that the effects of some scalar associated with flavor
violation affects the running of the Yukawa couplings in such a way,
as the renormalization group scale crosses its mass threshold.

One of the hints that the $Y_d\sim Y_u Y_e^\dagger$ relation might be
correct is that it naturally leads to large mixing angles in the
leptonic sector.  In this paper we have suggested a completion of the
framework that includes the neutrino Yukawa matrix, such that
$Y_\nu \sim Y_e Y_u^\dagger$.  This is not as predictive as the
original relation, because it does not specify the structure of the
heavy neutrino mass matrix.  The latter we have fixed (up to phases)
using experimental constraints on neutrino masses and mixings.

As an application, we studied leptogenesis within this framework,
whose heavy neutrino masses are very hierarchical.   It was found to
give an example where decays of the intermediate heavy neutrino $N_2$
give the dominant contribution to the baryon asymmetry.  In a random
scan, the models with the highest probability of giving large enough
asymmetry are those with normal mass hierarchy for the light
neutrinos, and mass $m_{\nu_1}\sim 0.01\,$eV for the lightest state.
This could be interpreted as a loose prediction of the model. It is an
interesting mass from the point of view of neutrinoless double beta
decay searches, since for $m_{\nu_1}\sim 0.01\,$eV, there is still a
reasonable chance of being sensitive to the effective  $|\langle
m_\nu\rangle|$ measured in $0\nu\beta\beta$  coming from the normal
hierarchy, while being able to distinguish it from that predicted  in
the inverted hierarchy.

We also noted that the flavor symmetry of the CFV scenario allows for
vector and scalar leptoquarks, constrained respectively at the scales
of $2$ and $260$ TeV.  The former is clearly in an interesting range
for the Large Hadron Collider, where ATLAS  \cite{Aad:2015caa} and CMS
\cite{Khachatryan:2015qda} have set lower limits near 1 TeV for
leptoquark masses.  In CFV they are predicted to have equal couplings
to all three generations.
\bigskip

{\bf Acknowledgments.}
JC acknowledges support from NSERC, and thanks NBIA for its generous
hospitality during the completion of this work.
ADF gratefully acknowledge support from the CONACyT for the postdoctoral fellowship
(grant No.\
237447) and to the department of physics at McGill University for its hospitality.
JR is supported in part by the International Postdoctoral Exchange
Fellowship Program of China.

\appendix

\section{Quark and neutrino masses and mixing angles}
\label{quark_masses}

In performing random scans to produce model realizations,  for the
quark sector, we use the quark masses and CKM matrix elements, along
with uncertainties, at the scale $m_Z$ as given by the Particle Data
Group \cite{PDG}.  For the models generated at the GUT scale, we use
the following central values and $1\sigma$ uncertainties
for quark masses from ref.\cite{Xing:2007fb},
in GeV:

\bea
	m_{u,c,t} &=& (0.48\pm 0.18)\times 10^{-3},\ 0.235\pm 0.04,\
74\pm 3.9 \nonumber\\
	m_{d,s,b} &=& (1.14\pm 0.5)\times 10^{-3},\ 22\pm 7\times
10^{-3},\ 1.00 \pm 0.04
	\nonumber\\
\eea
The uncertainties are derived assuming that $\Delta m({\rm GUT}) =
\Delta m(m_Z) \times
(m({\rm GUT})/m(m_Z))$.  For the CKM matrix, we take the PDG
values and uncertainties at scale $m_Z$, using the exactly unitary
parametrization (12.3-12.4) of
http://pdg.lbl.gov/2014/reviews/rpp2014-rev-ckm-matrix.pdf.  At the
GUT scale we take the central values of
\cite{Fusaoka:1998vc}.  In the Wolfenstein parametrization,
\bea
	\lambda &=& 0.22045\pm 0.00061, \quad A= 0.8797\pm 0.024,
	\nonumber\\
	\rho &=& 0.0\pm 0.021,\qquad\qquad\  \eta=0.371\pm0.013
\eea
still using the PDG errors from the $m_Z$ scale.

For the neutrino sector, we do parameter scans using the
$U_{\sss PMNS}$ mixing angles and neutrino mass differences
\cite{Gonzalez-Garcia:2014bfa},
{\small
\begin{eqnarray}
&&\sin^2\theta_{12}=0.304^{+0.013}_{-0.012},\nonumber\\
&&\sin^2\theta_{23}=0.452^{+0.052}_{-0.028},\nonumber\\
&&\sin^2\theta_{13}=0.0218^{+0.0010}_{-0.0010},\nonumber\\
&&\Delta m_{12}^2=7.50^{+0.19}_{-0.17}\times 10^{-5}\textrm{eV}^2,\nonumber\\
&&\Delta m_{31}^2=2.457^{+0.047}_{-0.047}\times 10^{-3}\textrm{eV}^2
\end{eqnarray}
}for normal hierarchy, and
{\small
\begin{eqnarray}
&&\sin^2\theta_{12}=0.304^{+0.013}_{-0.012},\nonumber\\
&&\sin^2\theta_{23}=0.579^{+0.025}_{-0.037},\nonumber\\
&&\sin^2\theta_{13}=0.0219^{+0.0011}_{-0.0010},\nonumber\\
&&\Delta m_{12}^2=7.50^{+0.19}_{-0.17}\times 10^{-5}\textrm{eV}^2,\nonumber\\
&&\Delta m_{32}^2=-2.449^{+0.048}_{-0.047}\times 10^{-3}\textrm{eV}^2
\end{eqnarray}
}for inverted hierarchy. These show the $1\sigma$ allowed regions,
while in our scans we vary to $3\sigma$.
The Dirac phase $\delta_{\textrm{CP}}$ is not constrained at
$3\sigma$.

\section{Efficiency factors}
\label{eff_fact}
Analytic fits to the efficiency factors quantifying washout of
the lepton asymmetries are given in ref.\ \cite{Buchmuller:2004nz},
for varying degrees of complexity in the initial conditions.
For the case of thermal leptogenesis, where the initial abundances of
heavy neutrinos is assumed to vanish for solving the Boltzmann
equations, the efficiency for decays of a given species with $K_i =
\tilde m_i/m_*$ can be expressed as $\kappa = \kappa^+ + \kappa^-$,
where
\be
	\kappa^+ = {2\over z}\left(1-e^{-\frac23 z \bar N}\right),
	\quad
	\kappa^- = -2 e^{-\frac23 N}\left(e^{\frac23 \bar N} - 1
	\right)
\ee
where
\bea
	z &=& K+\frac{K}{2}\ln\left(1 + {\pi K^2\over 1024}\ln\left[{3125\pi
	K^2\over 1024}\right]^5\right)\nonumber\\
	N &=& {9\pi\over 16}K,\quad
	\bar N = N\left(1 + \sqrt{\sfrac43 N}\right)^{-2}
\eea
Unlike eq.\ (\ref{Kieq}), here $\kappa$ vanishes in the weak washout
limit $K\to 0$, due to the initial condition of vanishing $N$
abundance at early times.


\end{document}